\documentclass{JINST}
\usepackage{epsfig}
\title{The PRO1 ASIC for Fast Wilkinson Encoding}
\author{L. L.~Ruckman\thanks{Corresponding author.}
$\; $ and G. S.~Varner\\
\llap{}Department of Physics and Astronomy, University of Hawaii,\\
  2505 Correa Road, Honolulu HI 96822, USA\\
  E-mail: \email{ruckman@hawaii.edu}}
\abstract{
Wilkinson conversion of stored samples in large Switch Capacitor
Array (SCA) ASICs, such as used for high speed waveform sampling, has many
benefits in terms of compactness, no missing output codes, low power
requirements and robustness.  However such Analog-to-Digital
conversions are relatively slow, limited by the encoder clock speed.
By repeating the same fast sampling technique used by the SCA, combined
with a fast priority encoder, significantly faster conversion is
demonstrated for a prototype ASIC designated PRO1.  For 8-10 bits of
resolution, this technique is compact and requires far fewer system
resources.}
\begin{document}

\section{Background}

Precision timing and amplitude instrumentation of large arrays of
photo detector elements for future applications in collider and
astroparticle physics has been enabled the proliferation of high-performance
and low-cost waveform sampling
devices~\cite{ATWD,Stefan,SAM,LAB3}.  To expand this technique
in a cost-effective manner to systems consisting of 0.1-1 million
channels, certain features could be quite useful.  In particular,
next-generation TeV gamma and Super B-factory detector applications
require trigger rates of 10's of kHz while providing multi-buffer
capability.  This requirement places a premium on analog conversion
performance.

\begin{figure}[ht]
\vspace*{0mm}
\centerline{\psfig{file=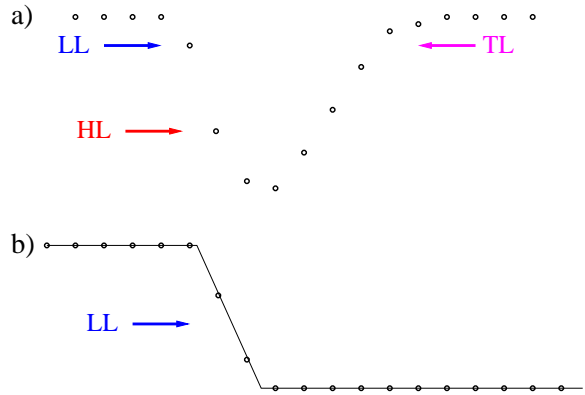,width=3.0in}}
\vspace*{0mm}
\caption{Flash encoding concept: a) is the analog waveform recording
with leading edge {\tt LL} and trailing edge {TL} for TOT and high
level {HL} for coarse Time Walk Correction; b) simplified scheme for
fast transition timing edge encoding.}
\label{PRO1_concept}
\end{figure}

We present the results of an ASIC developed for the flash encoding of
photodetector signals, a number of methods of which have been
evaluated~\cite{PROMPT}.  This concept is an outgrowth of earlier
work~\cite{JINST1} and is illustrated in Fig.~\ref{PRO1_concept} part
a) where the leading and trailing edge times are used to determine the
timing and Time-Over-Threshold (TOT) for an analog waveform.  A high
level threshold crossing may be used to improve the intrinsic time
determination error due to amplitude dependence (``time walk'').  Part
b) of the figure illustrates the flash time encoding of the digital
output of a ramp (Wilkinson) comparator.
\begin{figure}[htb]
\vspace*{0mm}
\centerline{\psfig{file=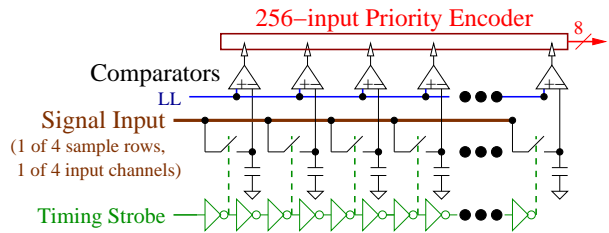,width=3.2in}}
\vspace*{0mm}
\caption{A block diagram of the PRO1 readout, where only the Low Level
({\tt LL}) threshold crossing output is considered.  A compact cascade
implementation for the priority encoder limited the settling time, and
can be improved.}
\label{BLOCK}
\end{figure}
\section{Architectural Details}

The Photodector Read Out version 1 (PRO1) ASIC was developed to
evaluate the use of waveform sampling in conjunction with threshold
crossing encoding to provide flash, although coarse, determination of
signal pulse parameters.  Relatively slow risetime signals, combined
with channel-channel comparator threshold spreads, resulted in limited
performance using this technique; at least for compact arrays and
precision timing (sub 100ps resolution) applications.  However for the
encoding of fast Wilkinson comparator outputs, the recording concept
illustrated in Fig. 1b) shows promise to improve upon limitations of
the high-speed Gray Code Counter (GCC) scheme usually
employed~\cite{BLAB1}.  In this fast Wilkinson technique, a ramp is
started coincident in time with the propagation of a write-pointer
strobe across the sampling array.  The comparator output transition
time is analog captured and each sample evaluated with a low-power
comparator.  The power required to operate each 8-bit sampling row is
about 9 mW, and could be lowered further by disabling the comparator bias
when conversion cycle is completed.

A priority encoder determines the location of the first threshold
crossing cell.  This signal flow is illustrated in Fig.~\ref{BLOCK}.
Typical effective times between samples in the array of 100's of ps
(many GSa/s effective) are common in 0.25-0.35$\mu$m CMOS
processes~\cite{ATWD,LAB3}.  Obtaining similar GHz digital counter
rates in a companion FPGA is either difficult or very power and
routing resource intensive.  

Table~\ref{FPGA_GCC} shows the 
system requirements for a Xilinx FPGA functioning as a Time-to-Digital 
Converter (TDC) using a high speed digital reference clock and a 
GCC.  A proper pipelined, dual clock phase GCC was used when 
simulating the amount of FPGA logic slices, flip flops, and Look-Up 
Tables (LUTs) required.  A proper pipelined GCC is a Gray Code counter 
that counts in Gray Code using an array of pipelined flip flops.  
Each PRO1 has the following specifications, relevant to application as
a TDC, as listed in Table~\ref{Specs}.

\begin{figure}[ht]
\vspace*{0mm}
\centerline{\psfig{file=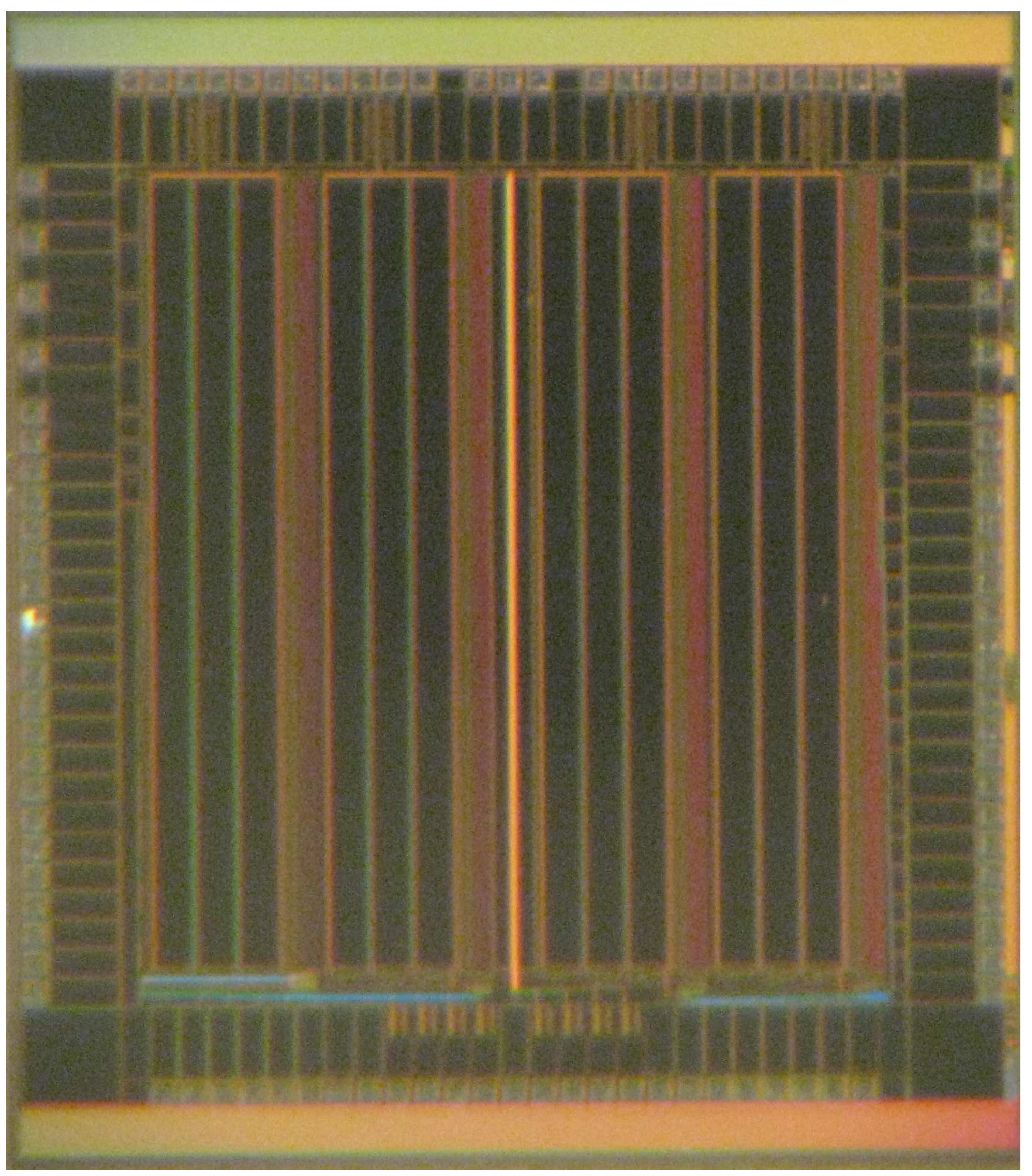,angle=270,width=3.0in}}
\vspace*{0mm}
\caption{A bare die photograph of the PRO1 ASIC.  The die is 3.21mm
by 3.03mm and is fabricated in the TSMC 0.25$\mu $m process.}
\label{bond}
\end{figure}

\begin{table}[hbt]
\caption{\it Programmable logic system requirements for a Xilinx Virtex or 
Spartan FPGA functioning as a TDC using a high speed digital reference clock
and a GCC.}
\label{FPGA_GCC}
 \begin{center}
    \begin{tabular}{|c|c|c|c|c|} \hline
      {\it \# of TDC CHs } & {\it GCC width (bits) } & {\it Logic Slices } & {\it Flip flops} & {\it LUTs}  \\ \hline\hline
      8  & 8  & 279  & 204  & 487  \\ \hline
      8  & 10 & 339  & 252  & 599  \\ \hline
      8  & 12 & 400  & 300  & 709  \\ \hline
      8  & 14 & 460  & 348  & 819  \\ \hline
      8  & 16 & 520  & 396  & 927  \\ \hline
      16 & 8  & 545  & 332  & 958  \\ \hline
      16 & 10 & 655  & 412  & 1166 \\ \hline
      16 & 12 & 764  & 492  & 1372 \\ \hline
      16 & 14 & 875  & 572  & 1578 \\ \hline
      16 & 16 & 983  & 652  & 1782 \\ \hline
      32 & 8  & 967  & 572  & 1742  \\ \hline
      32 & 10 & 1179 & 716  & 2142 \\ \hline
      32 & 12 & 1390 & 860  & 2540 \\ \hline
      32 & 14 & 1601 & 1004 & 2938 \\ \hline
      32 & 16 & 1813 & 1148 & 3334 \\ \hline
     \end{tabular}
  \end{center}
\end{table}   
\begin{table}[hbt]
\caption{\it Relevant sampling specification for the PRO1 ASIC when
used as a flash TDC.  Measurements from a single channel, single
storage row, consisting of 256 samples are presented.}
\label{Specs}
 \begin{center}
    \begin{tabular}{|l|c|l|} \hline
      {\it Parameter }& {\it Value } & {\it Unit}  \\ \hline\hline
      Sampling Rate  & 1.0 - 2.7 & GSa/s \\ \hline
      Full range  & 95-250 & ns \\ \hline
      Nominal Time Step &  400 & ps (2.5 GSa/s) \\ \hline
      Number of inputs &  4 & channels per PRO1 \\ \hline
      Sample rows & 4 & per channel \\ \hline
      Encoding output & 8 & bits  \\ \hline
     \end{tabular}
  \end{center}
\end{table}   
Fig.~\ref{bond} shows a photograph of the PRO1 bare die.  As noted
earlier, additional circuitry exists for prototyping other
functionality.  For the measurement reported, only a single channel
and storage row are considered, and only the Low-level comparator
output thereof.  When implementing only that functionality, the density
clearly can and will be increased, though a constraint is provided by
the need to reduce the settling time of the select logic tree
employed.  This logic can take as long as the write-pointer
propagation time across the array to settle (100's of ns).

\section{Readout Test System}

\begin{figure}[ht]
\vspace*{0mm}
\centerline{\psfig{file=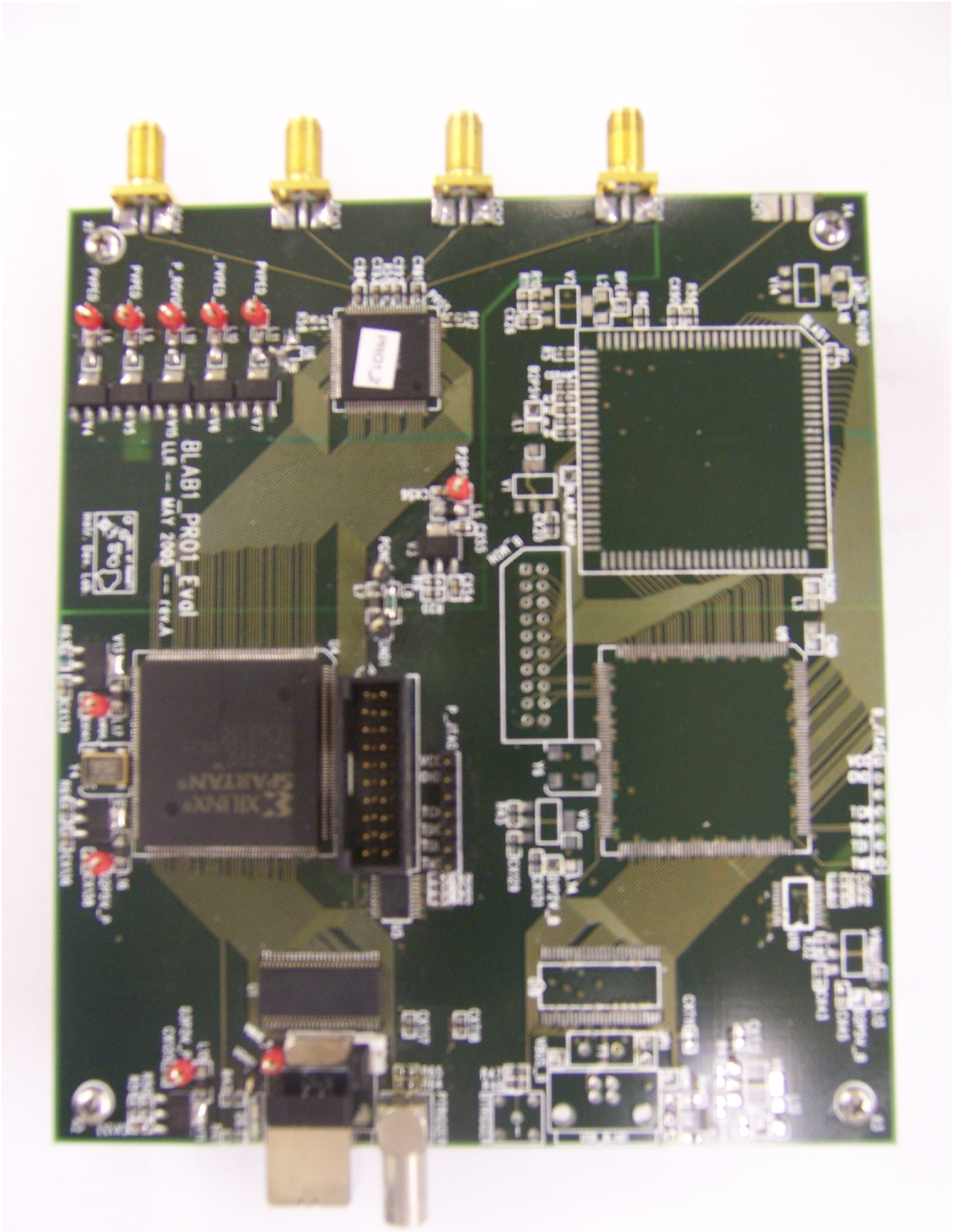,width=3.0in}}
\vspace*{0mm}
\caption{Photograph of the PRO1 evaluation circuit board}
\label{PRO1_eval_PCB}
\end{figure}

A printed circuit board was fabricated to evaluate PRO1 performance, a
photograph of which which is shown in Fig.~\ref{PRO1_eval_PCB}.  The
three main components on this circuit board are a packaged PRO1
ASIC, an FPGA, and a Universal Serial Bus (USB) interface chip.  The
external communication interface is via USB 2.0 and the Cypress
CY7C68013-56PVC USB microcontroller is used.  This USB microcontroller
controls the data being sent to and received from the FPGA to a
computer interface.  The FPGA controls the digital logic and timing
for the PRO1 readout, and the Xilinx XC3S200 is used.  RAM banks
internal to the FPGA buffer the digitized values while the data is
being dumped into the USB data stream.  A basic software tool was
developed to send commands to the FPGA and record PRO1 data via the USB
2.0 interface.


\section{Test Results}

By using the readout system described in the previous section, 
a number of the basic performance parameters of the
PRO1 ASIC were evaluated. Because timing performance is suchVer.1.2 2008/11/03 a
critical feature of this ASIC functioning as a TDC for Wilkinson conversion, 
each parameter is described in detail in the following subsequent sections.


\subsection{Sampling speed}

\begin{figure}[ht]
\vspace*{0mm}
\centerline{\psfig{file=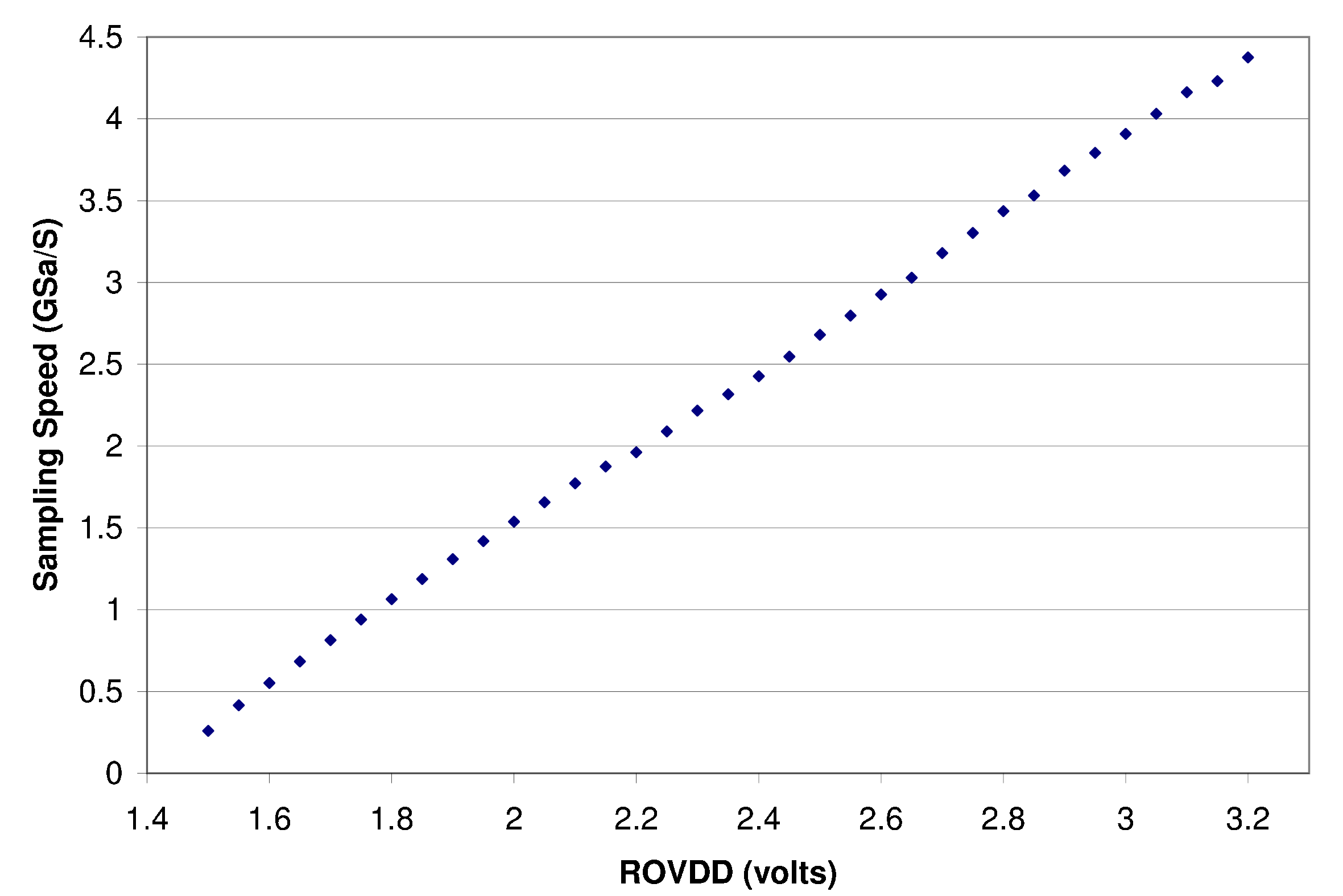,width=3.0in}}
\vspace*{0mm}
\caption{Sampling rate as a function of the ROVDD control
voltage, where extended operation ($\>$ 2.5V) is possible.}
\label{rovdd}
\end{figure}

Determination of the sampling speed is made by
measuring the time interval between insertion of
the timing strobe and appearance of the output
pulse from the last cell of the row, minus pad buffer
delays. The sampling speed is calculated by taking
the number of cells in a row and dividing it by
the propagation time for a given control voltage
setting. A plot of the sampling speed versus control
voltage (ROVDD) is shown in Fig.~\ref{rovdd}, where it is
seen that sampling rates from below 0.3 GSa/s to
above 4.5 GSa/s are possible.

 \subsection{Temperature Dependence}
\begin{figure}[ht]
\vspace*{0mm}
\centerline{\psfig{file=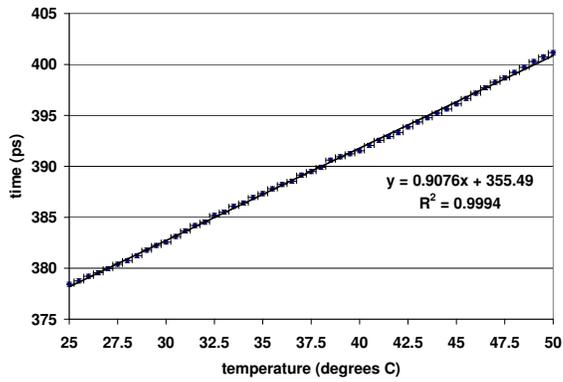,width=3.0in}}
\vspace*{0mm}
\caption{Temperature dependence of the sampling rate.}
\label{TEMP}
\end{figure}

One potential disadvantage of this voltage controlled delay technique
is that the circuit is temperature dependent. This dependence is seen
in Fig.~\ref{TEMP} and is roughly 0.3$\%/^\circ\mathrm{C}$, around
room temperature, and completely matches expectation from SPICE
simulation.  While for many applications, this variation would not be
significant and can be calibrated out with an external reference
clock~\cite{LAB3}.
\begin{figure}[ht]
\vspace*{0mm}
\centerline{\psfig{file=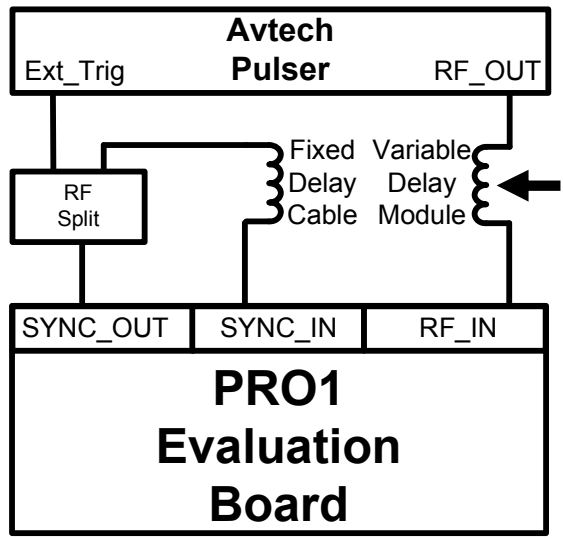,width=3.0in}}
\vspace*{0mm}
\caption{Schematic of the PRO1 timing measurement.}
\label{TIME_SETUP}
\end{figure}

\subsection{Timing Performance}

The PRO1 timing performance was evaluated using the test setup shown
in Fig.~\ref{TIME_SETUP}.  A synchronization pulse from the evaluation
circuit board goes through an RF splitter.  One copy of the sync pulse
is fed back with fixed cable delay into the evaluation board to create
the sampling strobe.  The other copy is used to trigger a Avtech
pulser.  The pulse from Avtech pulser was tuned to a 500 mV amplitude with a 
0.5 ns rise time and 10 ns full width half max (FWHM) duration. The discriminator 
on the PRO1 ASIC was set to trigger on the pulse's rising edge at 250 mV.  
The Avtech AVMP-2-C-P-EPIA pulser was used and its output
was inserted into the RF input of the PRO1 ASIC and scanned across the
sampling window using a variable delay module.

\begin{figure}[ht]
\vspace*{0mm}
\centerline{\psfig{file=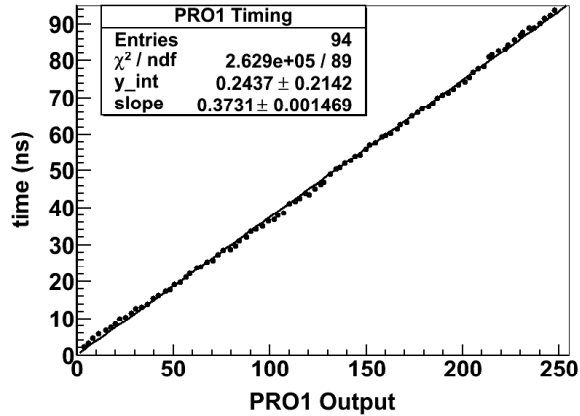,width=3.0in}}
\vspace*{0mm}
\caption{Plot showing the linear response of the PRO1 output 
with respect to a fixed time displacement.}
\label{LINEAR}
\end{figure}
\begin{figure}[ht]
\vspace*{0mm}
\centerline{\psfig{file=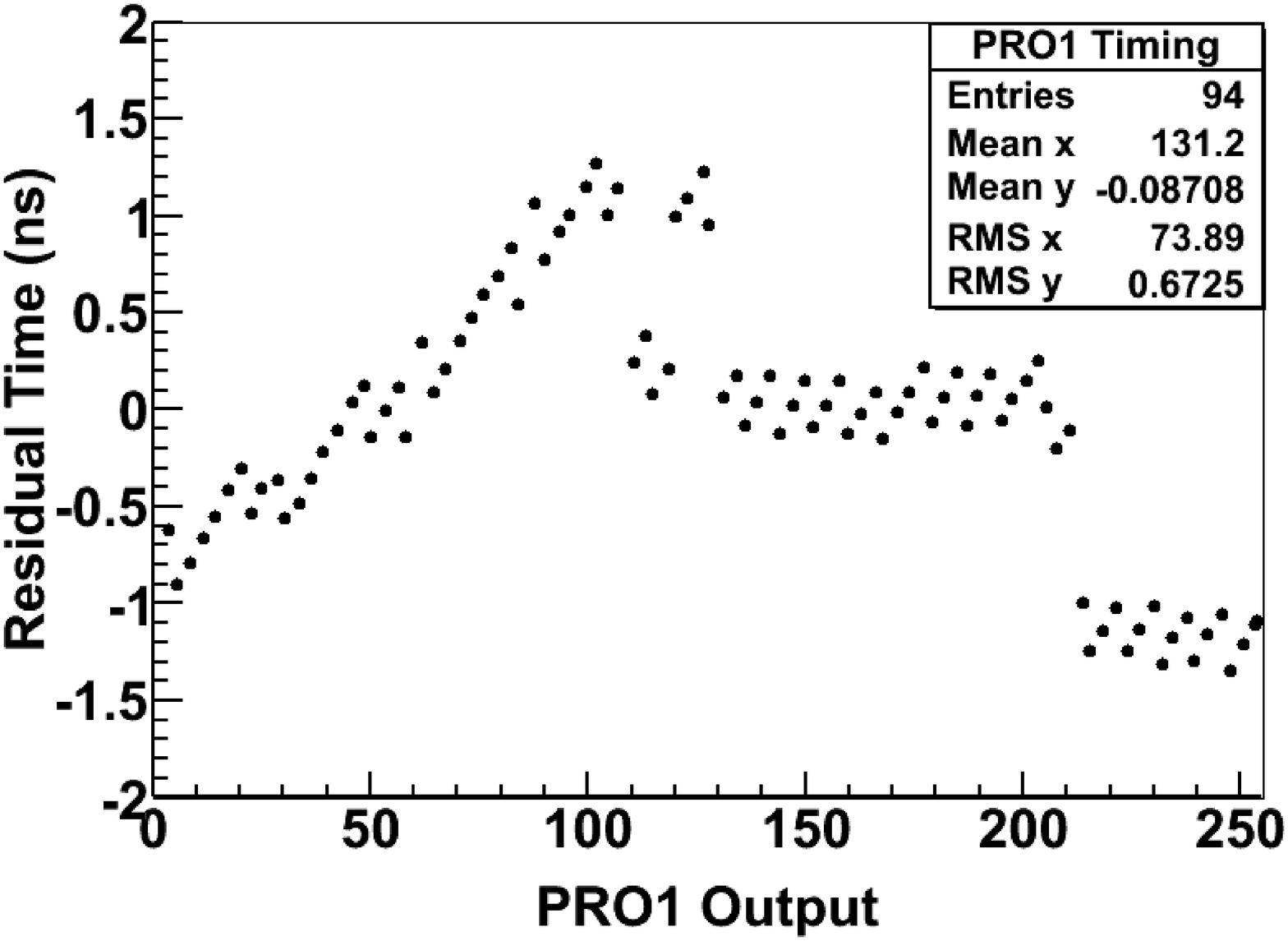,width=3.0in}}
\vspace*{0mm}
\caption{Plot of the residual data structure from 
subtracting the linear fit to the data points.}
\label{RESIDUAL}
\end{figure}
\begin{figure}[ht]
\vspace*{0mm}
\centerline{\psfig{file=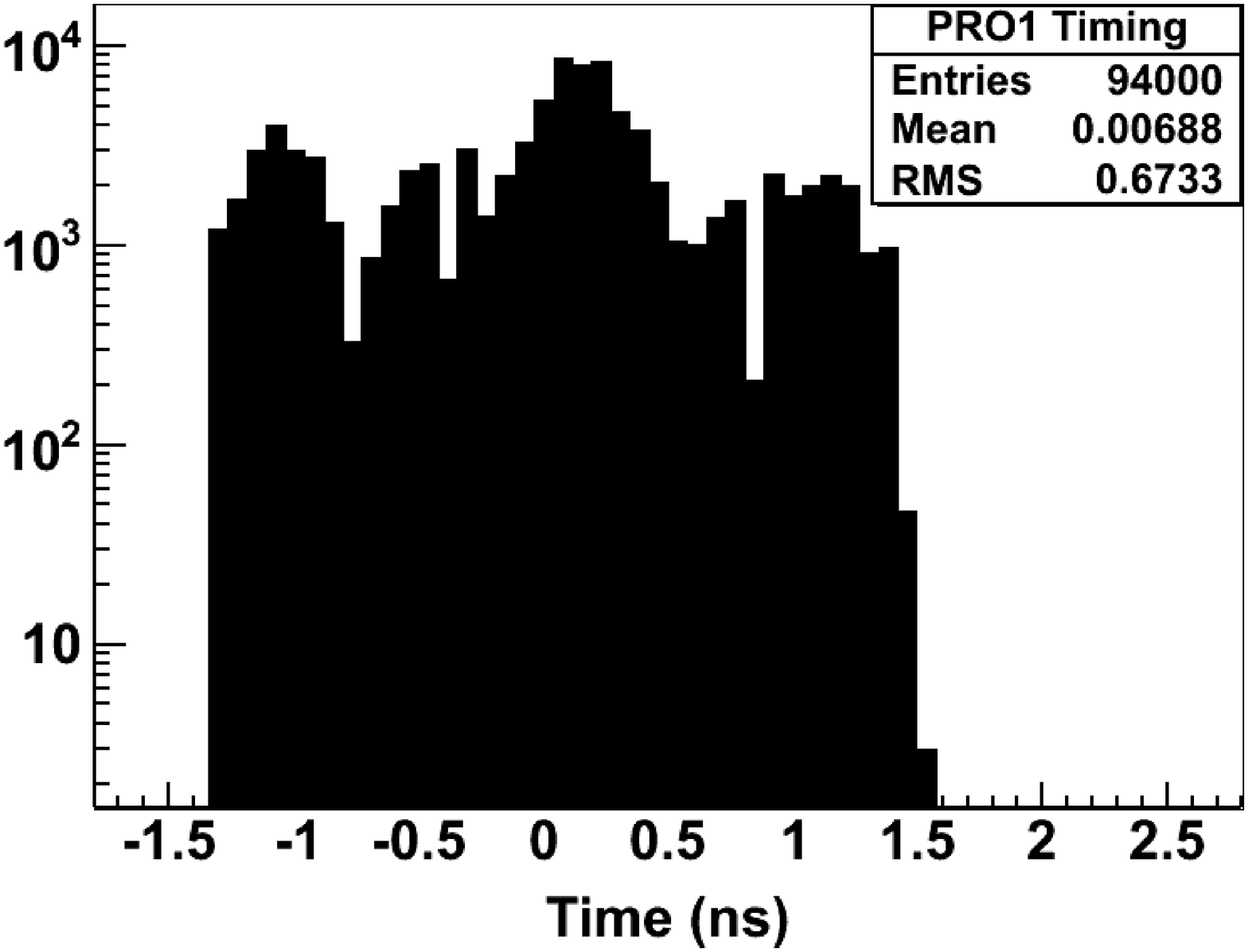,width=3.0in}}
\vspace*{0mm}
\caption{Histogram showing the timing jitter with no calibrations.}
\label{TIME_NO_CORR}
\end{figure}
\begin{figure}[ht]
\vspace*{0mm}
\centerline{\psfig{file=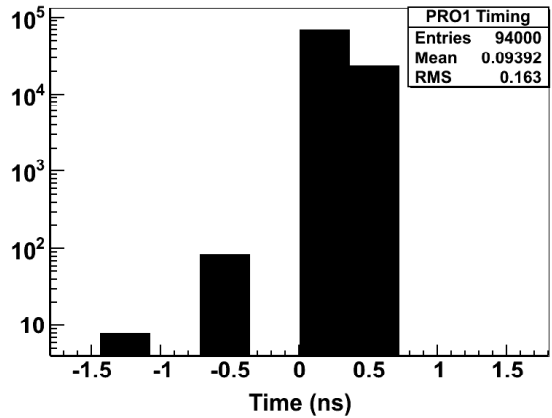,width=3.0in}}
\vspace*{0mm}
\caption{Histogram showing the timing jitter with calibrations.}
\label{TIME_CORR}
\end{figure}

By scanning the sampling window with this test setup, the linear
response of the ASIC's digital output versus the time difference is
shown in Fig.~\ref{LINEAR}.  Each data point for Fig.~\ref{LINEAR} is
the average PRO1 output for 10k events.  The slope of the linear fit
shows that the average time step to be 373 ps between sampling pixels
with ROVDD tied to the ASIC's VDD.  The substraction of the linear fit
to the data points is shown in Fig.~\ref{RESIDUAL}.  The structure in
the residual plot can be used to create a bin-by-bin correction to
improve PRO1 ASIC timing.  The projection of the all residual events
is shown in Fig.~\ref{TIME_NO_CORR}, which has an RMS timing jitter of
about 673ps.  By applying a bin-by-bin timing calibration to the PRO1
ASIC's digital output, the RMS timing jitter is reduced down to 163ps,
which is shown in Fig.~\ref{TIME_CORR}.  This TDC performance, after
applying the calibration corrections, is not so far from the ideal
binary interpolation $\frac{1}{\sqrt{12}}$ limit (108ps).  The additional
timing error is attributed to sampling speed temperature
drift and storage cell dependent comparator threshold dispersion.


\section{ADC Implementation}

By using this circuit for Wilkinson conversion, a calibration must be 
performed to remove the systematic errors.  This calibration procedure is 
done by applying a fixed DC voltage to the analog input of the Wilkinson ADC.  
By stepping through different fixed DC voltages within the ADC digitizing 
dynamic range, the average PRO1 output is mapped as a function of voltage.  
From these calibration measurements, a look-up table can be generated to 
convert the PRO1 output into voltage for a Wilkinson conversion application.

The method of using time interpolation by digital delay lines for 
boosted Wilkinson conversion has been demonstrated to achieve TDC resolution 
as low as 20 ps~\cite{DDL}.  While the PRO1 TDC resolution is roughly 20 times 
courser and slower than the digital delay lines method, 
the PRO1 method is useful in applications with 
a fast readout duty cycle using deep SCA storage for buffering during the readout 
deadtime.  Since the PRO1 TDC method doesn't require an 
external clock reference, the stored analog 
voltage on the SCA has reduced coupling of switching noise from the absents 
of a reference TDC clock.

\section{Future Directions}

Demonstration of the efficacy of this technique has led to its
adoption in the 2nd generation of large, buffered analog storage
device for precision timing ASIC (BLAB2).  It will also be featured in
a device intendend for continuous monitoring of turn-by-turn x-ray
emission of high luminosity electron storage rings (STURM).  In both
cases the low-power and faster conversion speed improve the density
and reduce processing speed and overall readout system overhead,
essential for future mega-channel readout systems.


\section{Summary}

A first generation of fast Wilkinson encoder CMOS device has been
studied in a 0.25 $\mu$m process.  This architecture is optimized to
reduce deadtime and power consumption while operating at an effective
multi-GHz digital counter rate for fast Wilkinson conversion.
Demonstrated low-power and high timing resolution makes this
architecture ideal for integrating a data collection FPGA with a SCA
waveform sampling ASIC, while reducing the amount of FPGA resources
needed.


\section{Acknowledgements}
Testing was supported in
part by Department of Energy Advanced Detector Research Award
\# DE-FG02-06ER41424.


\end{document}